\documentclass[letterpaper]{EslabStyle}
\usepackage{graphicx}

\newcommand{\solarmass}{${\rm M}_{\sun}$~} 
\newcommand{\solarmassyr}{\solarmass ${\rm yr}^{-1}$~} 
 
\newcommand{\Mdot}{$\dot{\rm M}$~} 
\newcommand{\Mstar}{${\rm M_{*}}$~} 
\newcommand{\Xin}{$X_{in}$~} 
\newcommand{\Xout}{$X_{out}$~} 

\begin{document}

\title{Water emission in NGC1333-IRAS4: Probing its surrounding envelope}
  
\titlerunning{Water emission in NGC1333-IRAS4} 
 
\author{S. Maret\inst{1} \and C. Ceccarelli\inst{2,3} \and 
   E. Caux\inst{1} \and A.G.G.M. Tielens\inst{4} \and 
   A. Castets\inst{2}} 
 
\institute{Centre d'Etude Spatiale des Rayonnements, CESR/CNRS-UPS, BP 
    4346, F-31028 Toulouse Cedex 04, France \and Observatoire de 
    Bordeaux, BP 89, F-33270 Floirac, France \and Laboratoire 
    d'Astrophysique, Observatoire de Grenoble, B.P. 53, F-38041 
    Grenoble Cedex 09, France \and Space Research Organization of the 
    Netherlands, P.O. Box 800, 9700 AV Groningen, The Netherlands} 

\authorrunning{Maret, Ceccarelli, Caux, et al.}

\titlerunning{Water emission in NGC1333-IRAS4} 

\maketitle 

\begin{abstract}
Sun like stars are born from the collapse of fragment of molecular
clouds. During the first embedded phase (the so called class 0/1), the
protostar is surrounded by a collapsing envelope, whose physical and
chemical structure would set up the initial conditions of the
proto-stellar disk which may eventually form planets. In this
contribution, we show how spectroscopic far infrared to millimeter
observations can be a powerful diagnostic to derive the chemical and
physical structure of the innermost regions of envelopes collapsing
around solar type protostars, where disk are believe to form. Our
claim is based on the simultaneous modeling of the dynamics, chemistry
and radiative transfer in a protostellar envelope. Here we present a
study of a class 0 protostar NGC1333-IRAS4, obtained using ISO-LWS
observations of water far-infrared lines. We show the derived physical
and chemical structure of the protostar, namely the density and
temperature profiles, as well as the water abundance across the
envelope. Besides, the mass of the central protostar and its accretion
rate can be constrained by these observations and modeling, yielding
an alternative method to measure these two poorly know parameters.


\end{abstract}

\section{Introduction}

NGC1333-IRAS4 is a binary protostellar system.  located in the south
part of the Perseus cloud. The relatively large mass of the
surrounding envelopes, together with the relative submillimeter to
bolometric luminosity ratio suggest that both source are highly
embedded and probably very young. They have been classified as
\emph{class 0} protostars (Andr\'e et al. 1993). IRAS4 is associated
with two molecular outflows, observed in the CO, CS (Blake et
al. 1995) and millimeters transitions (Lefloch et al. 1998). The
dynamical age of these outflow is a few thousand of years. In the past
years, several studies have focused on the molecular emission of this
source. In particular, Blake et al. carried out a study of several
molecules in NGC1333-IRAS4. Their major results were large depletion
factors of all molecules in the envelopes, and the presence of a
region with an increased abundance of CS, SiO and CH$_3$OH that the
authors attribute to grain mantle desorption caused by grain-grain
collisions induced by the powerful outflow originating from this
source. More recently, Di Francesco et al. (2001) reported the
detection of inverse P-Cygni profiles of several lines toward
NGC1333-IRAS4A, providing the least ambiguous detection of infall
motion towards a protostar ever. Here we present a study of the far
infrared water line spectrum of NGC1333-IRAS4 observed with the
\emph{Long Wavelength Spectrometer} (Clegg et al. 1996, herein after
LWS) on board ISO (Kessler et al. 1996). Using an accurate model of
the thermal emission of the envelope, we show how the physical and
chemical structure of the protostar can be derived (see also Maret et
al. 2002).

\section{Observations}

A full range spectral survey of (43-196 $\mu$m) of NGC1333-IRAS4 was
performed using LWS in the low resolution (R $\sim 200$) grating
mode. The observations were obtained on three positions, the first one
beeing centered between NGC1333-IRAS4A and NGC1333-IRAS4B, and the two
other were taken along the outflow powered by NGC1333-IRAS4, NW and SW
of it respectively. The Fig. 1 shows the spectra observed on the tree
positions. We detected fourteen H$_2$O lines, seven CO lines, three OH
lines, the two [OI] and one [CII] fine structure line towards the
central position. No significant H$_2$O, CO and OH emission was
detected on the NE and SW positions, whereas the [OI] and [CII] lines
were detected towards the three position at the same intensity

\begin{figure}[ht]
\resizebox{\hsize}{!}{\includegraphics{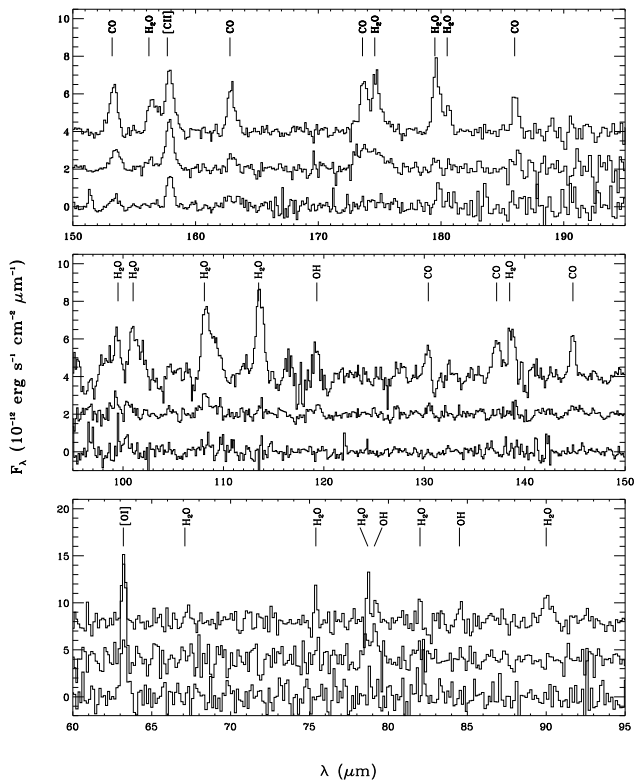}}
\caption{ISO-LWS spectra obtained towards NGC1333-IRAS4A on the
  central (top line), NE (middle line) and SW (bottom lines) positions.}
\end{figure}

\section{Line modeling}

The observed emission observed towards NGC1333-IRAS4 can have a priori
three different origin: the outflow powered by this source, the PDR at
the surface of the cloud, or the thermal emission of the protostellar
envelopes. Origin of molecular emission can be usually determined if a
sufficient enough spatial or spectral resolution is achievable. In the
case of ISO-LWS observations, all the gas components are mixed in the
80'' beam of ISO. Moreover, the relatively low spectral resolution
does not allow to spectrally disentangle these components. The
outflows powered by protostars has been invocated in the literature as
a possible origin of the water line emission observed towards them
(e.g. Giannini et al. 2001), but the absence of detection of water
emission on the outflow lobes does not play in favor of this
hypothesis. Moreover, Ceccarelli et al. (1999) noted the absence of
the correlation between the H$_2$O and SiO emission\footnote{SiO is
usually associated with the outflow strong shocks (e.g. Bachiller \&
Perez Gutierrez 1997) and it is believed to be a product of grain
mantle desorption (Caselli et al. 1997; Schilke et al. 1997)} toward
seven protostars, whereas they noted a correlation between the 1.3 mm
envelope emission. This suggest that the FIR water line emission can
be due to the thermal emission of the envelopes surrounding the two
protostars.\\
In order to test this hypothesis, we modeled the water emission by an
accurate model which simultaneously compute the thermal balance,
chemistry and radiative transfer in the envelope (Ceccarelli,
Hollenbach \& Tielens 1996). The model has four free parameters: the
central mass of the protostar \Mstar, its accretion rate \Mdot, and
the water abundance across the envelope. In order to take into account
the increase of water abundance in the inner parts of the envelope --
due to the evaporation of grain mantle when the gas reach the
sublimation temperature of water ices -- we used a different abundance
in the inner (\Xin) and outer parts (\Xout) of the envelope. We run
several models, varying the central mass between 0.3 and 1 \solarmass,
the accretion rate between 10$^{-5}$ and 10$^{-4}$ \solarmassyr, and
the water abundance in the outer and inner parts between
10$^{-7}$--10$^{-6}$ and 10$^{-6}$--10$^{-5}$ respectively. The best
fit parameters were obtained by minimizing the $\chi^2$ between the
observations and the model results. The minimum $\chi^2$ value is
obtained for \Mstar = 0.5 \solarmass, \Mdot = 5 $\times$ 10$^{-5}$
\solarmassyr, \Xout = 5 $\times$ 10$^{-7}$ and \Xin = 5$\times$
10$^{-6}$. The model reproduce relatively well the observed water
emission, with the exception of two lines which seems to be
underestimated by our model.

\section{Discussion}

\begin{figure}[ht]
\resizebox{\hsize}{!}{\includegraphics{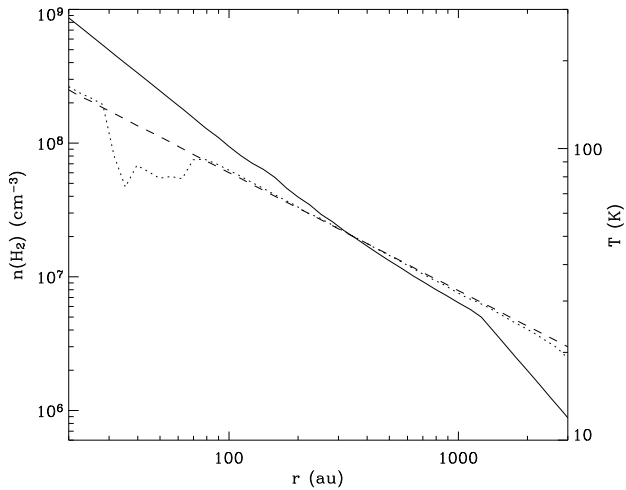}}
\caption{Density and temperature derived from the model. The solid
  line refer to the density profile in the envelope, the dashed line
  refers to the dust temperature, while the dotted line refers to the
  gas temperature.}
\end{figure}

The first result of this study is the capability of to reproduce the
observed water line emission with an envelope thermal emission
model. Our model also allowed us to derive the density and temperature
profile in the envelope. The Fig. 2 show the density and temperature
profile derived. The observations are consistent with a Shu (1977)
``inside-out'' density profile, with a transition region starting at
1500 au. This profile is in good agreement with the one derived by
J{\o}rgensen et al. (2002), using continuum observations. The dust
temperature closely track the dust one, with the exception of a small
region at R $\sim$ 50 au. This is causes at the radius where the
temperature reach 100 K, and the grain mantle evaporates, releasing
large amount of water in the gas phase (about a factor ten more),
increasing the cooling efficiency of the gas. This \emph{hot core like
region}, similar to the ones in massive star forming region -- here
indirectly probed by the jump in the water abundance needed to
reproduce the observed emission -- as been already observed on
IRAS16293-2422, where a similar study was performed (Ceccarelli et
al. 2000a, 2000b). This claimed was confirmed by a recent study by
Sch{\"o}ier et al. (2002). Furthermore, the central mass of the
protostar is also constrained by our model, giving an alternative
measure to this poorly known parameters. We derived a central mass of
0.5 \solarmass and an accretion rate of 5 $\times$ 10$^{-5}$
\solarmassyr. This gives an age of 10$^{4}$ year, close to the
dynamical age of the outflow. This values compares well with the
values derived recently by Di Francesco et al. The water abundance we
found for the outer envelope is in good agreement with recent SWAS
observations of the Perseus cloud (Bergin et al. 2002). As inner water
abundances is dominated by the grain mantle evaporation, the derived
inner abundance should be a indirect measure of the water abundance in
the grain mantle. the abundance we derive is somewhat lower than the
water ice abundance observed towards massive star forming region
(Tielens et al. 1984). Finally the abundances derived for
NGC1333-IRAS4 are in good agreement with the one derived by Ceccarelli
et al (2000a) for IRAS16293-2422. This study show that NGC1333-IRAS4
is probably younger than IRAS16293-2422.

\section{Conclusion}

Using a accurate model of the thermal emission of a protostellar
envelope, we successfully modeled the water emission observed with
ISO-LWS. We derived density and temperature profile in the envelopes,
as well as the water abundance. The water abundance is enhanced by a
factor ten in the inner parts, in an \emph{hot-core like} region. We
also derived the central mass of the protostar, which gives us an
estimate of its age, if the accretion rate remains constant. Both
mass, accretion rate, and water abundances compares well with recent
studies. This work also emphasis the need of ground based millimeter
observation, where higher spatial and spectral resolutions are
achievable. From this point of view, H$_2$CO and CH$_3$OH are of
particular interest because of their large abundances in grain
mantle. H$_2$CO and CH$_3$OH abundances should therefore be increased
in the central parts of envelopes. This is indeed confirmed by very
recent observations of these molecules towards several class 0
protostars (Maret et al. in prep.).

\end{document}